\documentclass[a4paper,fleqn]{cas-dc}

\usepackage{balance} 
\usepackage[authoryear]{natbib}
\usepackage[english]{babel}
\usepackage[utf8]{inputenc}
\usepackage[T1]{fontenc}
\usepackage{microtype}

\usepackage{amsmath}
\usepackage{amsthm}
\usepackage{algorithm}
\floatname{algorithm}{Operation}
\usepackage{algorithmic}
\usepackage{pifont}
\usepackage{graphicx}
\usepackage{makecell}

\newcommand{\engineer}[1]{\emph{``#1''}}
\newcommand{\related}{\textsc{related work} }
\newcommand{\findings}{\textsc{findings} }
\newcommand*\rot{\rotatebox{90}}

\newcommand*\OK{\ding{51}}

\def\tsc#1{\csdef{#1}{\textsc{\lowercase{#1}}\xspace}}
\tsc{WGM}
\tsc{QE}
\tsc{EP}
\tsc{PMS}
\tsc{BEC}
\tsc{DE}

\graphicspath{{./}}
\DeclareGraphicsExtensions{.pdf,.png}

\begin{document}
\let\WriteBookmarks\relax
\def\floatpagepagefraction{1}
\def\textpagefraction{.001}
\shorttitle{An Empirical Characterization of Event Sourced Systems}
\shortauthors{Overeem et~al.}

\title[mode = title]{An Empirical Characterization of Event Sourced Systems and Their Schema Evolution - Lessons from Industry}     
\tnotemark[1]

\tnotetext[1]{This research was supported by the NWO AMUSE project (628.006.001): a collaboration between Vrije Universiteit Amsterdam, Utrecht University, and AFAS Software in the Netherlands.}

\author[afas,uu]{Michiel Overeem}[orcid=0000-0003-4807-4124]
\cormark[1]
\cortext[cor1]{Corresponding author}
\ead{michiel.overeem@afas.nl}
\author[afas]{Marten Spoor}
\ead{marten.spoor@afas.nl}
\author[uu,lut]{Slinger Jansen}[orcid=0000-0003-3752-2868]
\ead{slinger.jansen@uu.nl}
\author[uu]{Sjaak Brinkkemper}[orcid=0000-0002-2977-8911]
\ead{s.brinkkemper@uu.nl}

\address[afas]{AFAS Software, Inspiratielaan 1, Leusden, The Netherlands}
\address[uu]{Utrecht University, Princetonplein 5, Utrecht, The Netherlands}
\address[lut]{Visiting Scientist, School of Engineering Science, LUT University, Finland}

\begin{abstract}
Event sourced systems are increasing in popularity because they are reliable, flexible, and scalable.
In this article, we point a microscope at a software architecture pattern that is rapidly gaining popularity in industry, but has not received as much attention from the scientific community. 
We do so through constructivist grounded theory, which proves a suitable qualitative method for extracting architectural knowledge from practitioners. \newline
Based on the discussion of 19 event sourced systems we explore the rationale for and the context of the event sourcing pattern.
A description of the pattern itself and its relation to other patterns as discussed with practitioners is given.
The description itself is grounded in the experience of 25 engineers, making it a reliable source for both new practitioners and scientists.
We identify five challenges that practitioners experience: event system evolution, the steep learning curve, lack of available technology, rebuilding projections, and data privacy.
For the first challenge of event system evolution, we uncover five tactics and solutions that support practitioners in their design choices when developing evolving event sourced systems: versioned events, weak schema, upcasting, in-place transformation, and copy-and-transform.

\end{abstract}


\maketitle

\section{Introduction}
\label{sec:introduction}

Software systems are increasing in complexity, used in increasingly critical processes, and serve increasing numbers of end-users. 
Architectural patterns enable engineers to built these systems using knowledge acquired by other engineers. 
Influential books such as \emph{Patterns of Enterprise Application Architecture} by~\cite{Fowler2002} and \emph{Enterprise Integration Patterns} by~\cite{Hohpe2004} demonstrate the impact of pattern descriptions on software engineering.
Architectural patterns are part of the trend of knowledge based architecture design; \citet{Li2013}.
\citet{Kassab2018},~\citet{Taibi2018a}, and~\citet{Harrison2007} show how patterns are instrumental in the capturing of architectural design decisions.
In this article, we describe such a pattern in detail and provide the design decisions that were employed in practice, with the goal of providing a comprehensive source of knowledge for practitioners.

Recently, the event sourcing pattern has become a popular answer to the challenges of complex, mission-critical, scalable systems.
Examples of organizations that apply event sourcing are Netflix~\citep{Avery2017}, and Walmart's Jet.com~\citep{Gorodinski2017}, with the goal of creating scalable and reliable critical systems.
Event sourcing is informally described by~\cite{Fowler2005} as a pattern that ``ensures that all changes to application state are stored as a sequence of events.''
Flexibility, debug-ability, and reliability are given by~\cite{Avery2017} as rationale for using event sourcing.
\cite{Debski2017} and \cite{Erb2016} show how event sourcing can be applied to achieve scalable, reactive systems.
\cite{Kabbedijk2013} describes event sourcing as a sub pattern of Command Query Responsibility Segregation (CQRS) in his work on the improved variability and scalability of systems applying CQRS.

 The events in event sourcing, as opposed to general event-driven architectures (EDAs)~\citep{Fowler2017}, are stored as an append-only log of all state changes.
Two key characteristics separate event sourcing from event-driven approaches, such as stream processing, transactional processing, and blockchain.
First, events in Event Sourced Systems (ESSs) are stored as the state of the application. 

Other approaches use the events to communicate, while the communication aspect comes second in ESSs. 
The second difference is that events are closely related to events occurring in real world business processes. 
This allows event sourcing to be also used as a design approach.
Domain-Driven Design~(DDD), as described by~\cite{Evans2003}, advocates events as a design tool for the process flow of a software system.
\cite{Brandolini2018} proposes \emph{event storming} (analogous to brainstorming), a group design process that focuses on the events that take place in a software system.
Further details on these analogous approaches are found in Section~\ref{sec:background}. 

Although event sourcing is related to existing ideas such as EDAs, the pattern itself has not yet been thoroughly studied.
Most knowledge exists in so-called `grey literature': practitioner blogs, and anecdotal experience reports.
In previous work~\citep{Overeem2017b}, that focused on the evolution of ESSs, we experienced this lack of literature.
This work fills this gap by deriving an integral description of event sourced systems through interviews with 25 engineers.
Together with this description we identify four categories of rationale for the application of event sourcing, such as a decrease of complexity. 
In this ``In Practice'' submission, we also identify five engineering challenges around the pattern, with schema evolution being one of the most complex challenges.
With the pattern description and its liabilities presented in this article, we enable engineers to make a considered choice. 
Our work is not dissimilar to the work of~\cite{Musil2015}, who conducted an extensive study on collective intelligence system pattern variations, with the goal of enabling architects to predict the outcomes of different design decisions. 
Similarly,~\cite{Slotos2016} describes the Star pattern for enabling flexible business applications, also with the goal of supporting software architecture researchers and practitioners and promoting the pattern itself.

Our study regards a new research area, therefore, we apply Grounded Theory (GT).
\citet{Adolph2011} describe GT as a useful approach for research in areas that have not previously been studied. 

A GT explains how people resolve their main concern by employing a certain process.
That process is called the `core category' of the GT. 
The core category of the work presented in this article is the process of designing and implementing event sourced systems, as performed by software engineers.
The theoretical definition of event sourcing helps both researchers and practitioners to understand, reason about, and teach the pattern and its consequences.
Section~\ref{sec:research-approach} explains how we applied GT to form a basis for conceptualization of ESSs from 25 interviews, and how the three essential elements are covered.
From the gathered data we distill the pattern description and its consequences.
This work has the following contributions:

\begin{itemize}
    \item Section~\ref{sec:background} contrasts ESSs with other existing architectural patterns, such as EDAs and blockchain, and shows that ESSs are insufficiently described in existing literature.
    \item Section~\ref{sec:esinpractice} describes the rationale for using ESSs: they provide audit functionality, are highly flexible and scalable, enable the development of highly complex systems, and are a current trend. 
    The overview of 19 different ESSs elaborates on the context of the pattern, showing that event sourcing is applied in different kinds of systems, from small to extremely large.
    \item Section~\ref{sec:eventstoreandsystem} provides a thorough description of ESSs based on the findings of the interviews, presenting the pattern itself including its relation to CQRS.
    It also reflects on the role of the (implicit) schema present in ESSs. 
    \item Section~\ref{sec:challenges} presents the \textbf{engineering challenges} surrounding the use of the pattern, that engineers encounter during the development of ESSs, such as a steep learning curve, poor ESSs performance, and dealing with privacy regulations such as the General Data Protection Regulation (GDPR).  
    \item Section~\ref{sec:evolution} focuses on of the most prominent challenge encountered in ESSs: \textbf{schema evolution.}
    Five empirically established methods are presented that support ESS evolution. 
    We advise that systems should start out using versioned events and weak schema, while later evolving to upcasting and even copy-and-transform techniques. 
\end{itemize}

The validity threats of this work, such as the fact that the interviewees were pragmatically collected, are discussed in Section~\ref{sec:threats}. 
We conclude that ESSs enable complex scalable systems with auditing capabilities and that our theoretical definition enables further research and development of these systems.

\section{Research Approach: Constructivist Grounded Theory}
\label{sec:research-approach}

In our early literature search, we identified that there is little academic material available when it comes to the topic of event sourcing.
Grounded Theory (GT) is defined as a systematic methodology involving the construction of theories through methodical gathering and analysis of data.
\cite{Adolph2011} explain how GT is particularly useful for research in areas that have not been studied before.
Our investigation of ESSs has an exploratory nature, therefore, we use GT to structure our research approach.
Furthermore, we aim to inspire researchers to experiment with novel approaches to gathering architecture knowledge. 

GT is a common research strategy in software engineering research and induces theory from empirically collected material, such as through interview or case studies. 
For instance,~\cite{Hoda2012} explore the practices of self-organizing agile teams using GT.
\cite{Greiler2012} apply GT to improve the understanding of testing practices for plug-in systems.
\cite{Tamburri2018} recover software architectures by applying GT.
Last, \cite{Santos2019} study common vulnerabilities in plug-and-play architectures through GT.

Similarly, we use GT to explore event sourcing, and improve our understanding of the pattern, the applications, and the challenges.
Constructivist GT assumes that neither data nor theories are discovered, but are constructed by the researchers out of the interactions with the field and its participants. Data are co-constructed by researchers and participants, and coloured by the researchers' perspectives, and values. 
Within this approach, a literature review is used in a constructive and data-sensitive way without forcing it on data.
We have employed constructivist GT~\citep{Charmaz1996} in our research; we knew we would find a description of the pattern, but were not aware what other concepts, challenges, and motivations would be identified.

\subsection{Research Questions and Motivation}
The motivation of our research is formed by five years experience in the development of an event sourced system and earlier research on schema evolution in ESSs~\citep{Overeem2017b}.
This experience guided our research and the direction of our exploration. 
Effectively, our previous work is also part of the GT data set, and has been translated directly into the research protocol.
The main goal of the research project was to come to a cohesive theory around the event sourcing architecture pattern. 
The research questions guided the research and were formulated, as per constructivist GT, a priori, but evolved to the following final set:

\begin{description}
\item[RQ1] What types of systems apply event sourcing and why?
\item[RQ2] How can event sourced systems be defined?
\item[RQ3] How can event sourced data structures be evolved?
\item[RQ4] What are the challenges faced by practitioners in applying event sourcing? 
\end{description}

Our previous study in the domain~\citep{Overeem2017b} gained significant industry interest, which led us to attend many industry events, where we were often invited as keynote speakers. 
This provided us extensive access to practitioners in the field, who would offer their support and advice. 
Through these rich interactions it became obvious that an extensive interview study could lead to new results and research challenges in the domain. 

\textbf{Foundations for the Study.} While in GT it is recommended that the researchers do not perform an extensive literature study before the research project, many have acknowledged that this is almost impossible and at times even impractical~\citep{Stol2015, Charmaz1996}. 
As little academic literature was available, it was easy to fulfill this major GT guideline.
This research project was started after we had already published in this domain~\citep{Overeem2017b} ourselves.
We made our previous work part of the initial data set and also included the works of Fowler, e.g.~\citep{Fowler2017}. 
The main concepts were extracted from these works and subsequently used to create an interview protocol.
Throughout the project, as we gathered new evidence and encountered new concepts, we performed exploratory literature study projects for each. 
Furthermore, if the interviewees mentioned an academic paper, it became part of our literature set. 
New concepts were extracted from this literature and integrated with the interview protocol where necessary. 
The literature was explored by snowballing forward and backward one level. 

\subsection{Sampling and Interviewees}
The interviewed engineers volunteered to contribute to our research after being invited through different channels.
Based on our experience in developing ESSs in the past years we identified the primary locations through which the event sourcing and DDD community communicates.
We invited the engineers through channels such as Google Groups and Slack channels.
In addition to this open invitation, we explicitly contacted and invited a number of well known community members.
We executed \emph{interview snowballing}, a process similar to snowballing in systematic literature studies~\cite{Wohlin2014}: we explicitly asked every interviewee for further references.
The interviewees were not compensated for their cooperation.

Our direct and indirect invitations resulted in interviews with 25 engineers.
The engineers are event sourcing practitioners in the roles of developers, architects, and product owners.
 A number of these engineers were consulting with the company, while others were employed by the company.
The consultants operate as external advisers (in addition to being hired as developer or architect) and are hired by multiple companies because of their experience. 
Table~\ref{tab:engineers} summarizes the engineers, including their role, years of experience with ESSs, the number of ESSs they worked on.
Combined they have 103 years of experience, with an average of four years per engineer.
For two of the engineers (E14, E16) it is hard to tell how many systems they worked on over the years, because their consultancy work exposed them to many different systems.
A number of the engineers worked on the same system(s), and were interviewed together.
 We conducted 22 distinct interviews with the 25 engineers.
Three interviews were conducted with two engineers together as these engineers worked on the same system.
In the case of E4 and E5, and E20 and E21 the engineers had a different role, and their experiences complemented each other during the interview.
Engineers E9 and E10 shared their role, and their answers showed more overlap. 
The systems are discussed in Section~\ref{sec:esinpractice}.
We will refer to the engineers by the number given to them in Table~\ref{tab:engineers}.

\begin{table}
  \renewcommand{\arraystretch}{1.25}
  
  \caption{
    \small{
        Summary of the interviewed engineers.
        We list roles (all technical except one), location, years of experience with ESSs and number of ESSs worked on.
    }
  }
  \label{tab:engineers}
  \centering
  \footnotesize
  \begin{tabular}{p{0.4cm}|p{2.6cm}|p{1.8cm}|r|r} 
    & Role & Location & \parbox{1.2cm}{Experience (years)} & \parbox{.6cm}{Nr ESSs}
  \\
  \hline
  E1  & Architect, Developer & North America & 4 & 3 \\
  E2  & Developer & Europe        & 2 & 1 \\
  E3  & Developer & North America & 2 & 1 \\
  E4  & Architect & Europe        & 2 & 1 \\
  E5  & Developer & Europe        & 2 & 1 \\
  E6  & Architect, Developer & Asia          & 15 & 3 \\
  E7  & Architect, Developer & Europe        & 4 & 3 \\
  E8  & Consulting Developer & Europe        & 2 & 1 \\
  E9  & Consulting Developer & Europe        & 3 & 2 \\
  E10 & Consulting Developer & Europe        & 3 & 2 \\
  E11 & Architect, Developer & North America & 9 & 3 \\
  E12 & Developer & Europe        & 3 & 1 \\
  E13 & Developer & Europe        & 2 & 1 \\
  E14 & Consulting Architect & Europe        & 10 & \emph{multi} \\
  E15 & Developer & Europe        & 1 & 1 \\
  E16 & Consulting Architect, \newline Developer & Europe        & 7 & \emph{multi} \\
  E17 & Architect & Europe        & 2 & 1 \\
  E18 & Architect & Europe        & 2 & 1 \\
  E19 & Architect & North America & 3 & 1 \\
  E20 & Product Manager & Europe        & 2 & 1 \\
  E21 & Architect & Europe        & 2 & 1 \\
  E22 & Architect & Asia          & 5 & 1 \\
  E23 & Architect & Europe        & 9 & 1 \\
  E24 & Architect & Asia          & 5 & 3 \\
  E25 & Developer & Europe        & 2 & 1
\end{tabular}

\end{table}

\subsection{Interview Techniques and GT}
Each interview took 30-90 minutes, either in person or via video conference. 
The protocol presented in the Appendix~\ref{sec:interviewprotocol} was created using the guidelines of~\cite{Castillo-Montoya2016}.
During the interviews, we asked open-ended questions exploring event sourced systems.
The questions asked during the interviews were based on a protocol that is downloadable with the interview transcripts~\citep{Overeem2021}. 

The protocol was followed freely: the answers given by the engineers guided the interviews.
The four protocol parts remained stable over the interviews. 
Some of the interview questions were sharpened and added as the interviews progressed, a technique encouraged by practitioners of GT.
The protocol used at the last interview is presented in the appendix.
The first part of the interview focuses on the context of the event sourced system and the engineer: what are the characteristics of the system, and why is event sourcing applied.
Versioning of event sourced systems is discussed in the second part of the interviews, based on our experience in this topic we identified this as an important challenge.
The third part deals with the relation of event sourcing with CQRS, DDD, and other challenges.
Finally, we discuss whatever the engineers thinks should be discussed in relation to event sourcing.

\subsection{Coding, Analysis, and Creativity} 

Each interview transcript was analyzed, as part of the GT approach, through an open coding process.
The interviews were conducted by the first author, the transcripts were reviewed by another author after creation.
The first and second author performed the codification and categorization, while the third author validated and confirmed the steps.
The authors maintained a shared \textit{memo-ing} document where ideas and emerging concepts were noted for discussion with all co-authors. 
Disagreements in the codification and categorization were resolved through discussions among the authors until agreement was found, while older versions of concepts were maintained in the memo-ing document.
The coding process was both organic and methodical. 

We provide an example of the coding process. 
One of the concepts that was discussed extensively was that of auditing and the ability to have a change log for all events in the system. 
E2: \textit{it ``has saved the finger of blame from pointing at us so many times ... that bit is worth its weight in gold to me.''} E4, translated: \textit{``I would save the old version forever ... for if we end up in court.''} 
Many of the interviewees put equal emphasis of the role of the audit log. 
The paragraphs from the transcripts mentioning the audit log were first coded and linked to the concept \textit{audit}.
From those codes \textit{audit} emerged as one of the prevalent rationales behind the pattern. 
After further grouping the statements linked to \textit{audit} we added more detailed codes, particularly addressing specializations of this rationale such as \textit{customer service support} and \textit{regulations}.

This example explains how we started with highlighting important paragraphs and sentences in the transcripts.
Those highlights were coded with short summary sentences.
After that the sentences were grouped by linking them to codes: topics described by a few words.
From those codes we derived concepts, such as the before mentioned \textit{audit}, which was later related to the category \textit{rationale}.
During this process we iterated until we ended with simplified categories and concepts (also known as the \emph{parsimony} principle) that reflected the linked paragraphs.
This process was iterative and organically executed until the first and second authors agreed on the categories and concepts.

While we cannot claim that saturation was reached, this article is a presentation of the coherent concepts that emerged from the research.
The nature of our study is exploratory and the research questions are broad on purpose.
To reach saturation on such a large topic one would have to conduct, transcribe, and codify an impractical number of interviews.
Although saturation based on the codes and concepts was not reached, we are confident that the results we present represent the overall sentiment among practitioners.
While we always had the concepts of how to present an architecture pattern in the back of our minds, we decided to structure the presentation according to the results of the GT concepts and codes.
 The guidelines as, for instance, stated by~\citet{Gamma1995} on describing a pattern through the elements \emph{problem}, \emph{solution}, and \emph{consequences} were used during the memo sorting process to match our concepts, but not as a predefined framework in which our concepts were painstakingly framed. 
Section~\ref{sec:discussion} discusses the relation between our concepts and the guidelines of~\citet{Gamma1995}.
The categories, concepts, and codes found during the interviews are presented in Sections~\ref{sec:esinpractice}, \ref{sec:eventstoreandsystem}, \ref{sec:challenges}, and \ref{sec:evolution}.
Tables~\ref{tab:rationales}, \ref{tab:systems}, \ref{tab:systemsizes}, \ref{tab:implementation}, \ref{tab:challenges}, and \ref{tab:techniques-summary} summarizes the results. 

The interview protocol, the anonymized transcripts of the interviews, and the classification codes with links to the interviews are made available as data package~\citep{Overeem2021}. 

\section{Background}
\label{sec:background}

The foundational idea of event sourcing is the domain event as described by~\cite{Evans2015}.
His seminal book on Domain-Driven Design (DDD), however, does not mention the pattern.
\cite{Vernon2013} describes event sourcing only briefly in his book on the implementation of various DDD patterns.
\cite{Young2017}, as one of the original proposers of event sourcing, discusses the challenge of versioning ESSs.
Event sourcing is also discussed in the context of CQRS~\cite{Young2010}, a pattern strongly related to event sourcing.
Recent academic literature \citep{Erb2019, Zhong2019} shows an interest in applying event sourcing for research projects.

Three related areas and their differences with respect to ESSs are discussed: transactional processing and database systems, stream processing and EDAs, and blockchain.

Event sourcing is related to database systems techniques used for persistence guarantees and replication.
\cite{Gray1992} describe how transaction logs can be used to replicate state between database systems.
Every state change is recorded as a transaction, which is similar to event sourcing where every state change is recorded as an event.
\cite{Kleppmann2017} discusses event sourcing in the context of data-intensive applications, he relates the pattern to the \emph{change data capture approach}, often used in Extract-Transform-Load (or ETL) processes~\citep{Vassiliadis2009}.
ETL solutions are often used for creating data warehouses.
The primary difference between event sourcing and these techniques is that a transaction or a data change is a technical entity without relation to the real world, while an event in event sourcing resembles an event in the real world. 

Kleppmann also relates event sourcing to the \emph{chronical data model} described by~\cite{Jagadish1995}, 
Time series, as described by~\cite{Dreyer1994}, is another data model that deals with the temporal aspects of data.
Both techniques are only used as a data modelling technique, while event sourcing is a software architecture pattern.

Event sourcing also shares commonalities with stream processing~\citep{Wu2006}, applied in for instance Internet of Things (IoT) systems to process sensor events.
Events in IoT systems are often used to communicate between different (sub)-systems, and are not stored as the state of the system.
Also, the events represent technical events such as sensor data as opposed to real world business domain events.
Another closely related topic is Complex Event Processing (CEP) as described by~\cite{Luckham2011}.
In CEP the focus is on pattern recognition within a stream of events.
CEP itself could be applied in the processing components within an ESS, similar as the event calculus formalism.
Event calculus, as described by~\cite{Sadri1995}, is a logical language that represents the effects of events.
This language, however, cannot be used to describe event sourcing as an architectural pattern.
Simarly, process mining deals with the analysis of event logs from process-driven systems.
The work of~\cite{DeMurillas2015} shows the complexity of mining processes from systems that do not record historical data.
ESSs support process mining by default, which makes them suitable for enterprise systems.

\cite{Anh2018} describes another append-only data structure: blockchain.
While the data structure is similar to event sourcing, the goals of the two techniques are different.
A blockchain focuses on solving problems related to distribution, consensus, and trust, while event sourcing solves problems with history, temporal complexity, and audit trails.
The blockchain approach enforces the immutability of the data to solve its problems, while in event sourcing this immutability is self imposed.
Event sourced systems could be build using a blockchain solution.
However, the distribution and consensus features offered by blockchain do not improve the goals targetted by event sourcing.

\section{Event Sourcing In Practice}
\label{sec:esinpractice}

The 25 interviewed engineers have an accumulated experience of at least 35 event sourced systems (ESSs).
However, a number of those systems were either not yet in production, or the engineer could not recall enough details of the system.
Of the 35 systems, 19 ESSs were discussed in more detail and are summarized in Table~\ref{tab:systems}.
Still, the experts experience on all of these systems is reflected in the answers that they gave, and is thus reflected in the challenges, the definitions, and the schema evolution techniques.
The categories in this characterization are based on the interviews, and were selected based on the categorization of the concepts deduced from the interviews. 

\begin{table*}
  \renewcommand{\arraystretch}{1.25}
  \caption{
    \small{Characterization of the ESSs under study, including the technology platform, the rationale for event sourcing and the chosen degree of immutability. 
    The application of Domain-Driven Design (DDD), the Microservice Architecture (MSA) style, and Command Query Responsibility Segregation (CQRS) is also indicated.}
  }
  \label{tab:systems}
  \centering
  \footnotesize
  \begin{tabular}{p{1.8cm}|p{1.2cm}|p{2.8cm}|p{2.8cm}|p{1.4cm}|p{1.7cm}|ccc} 
     \textbf{System Code}
     & \textbf{Engineers}
     & \textbf{Type of application}
     & \textbf{Technology platform}
     & \parbox[b]{\hsize}{\textbf{Rationale}}
     & \parbox[b]{\hsize}{\textbf{Degree of \newline immutability}}
     & \rot{\textbf{DDD}} & \rot{\textbf{MSA}} & \rot{\textbf{CQRS}} \\
  \hline
  MarketingSys & E22
     & Marketing automation
     & .NET, DynamoDB
     & audit
     & strict
     & \OK & \OK & \OK
  \\ \hline
  HealthSys & E23
     & \parbox[t][][t]{\hsize}{Health record \newline management}
     & JVM, MySQL
     & \parbox[t][][t]{\hsize}{audit, \newline flexibility}
     & \parbox[t][][t]{\hsize}{cut-off \newline moments}
     & \OK &     & \OK
  \\ \hline
  WebBuildSys & E24
     & Website building
     & Scala, MySQL
     & audit
     & strict
     &     & \OK & \OK
  \\ \hline
  B2CSys & E1
     & B2C communication
     & JVM, MongoDB
     & flexibility
     & strict
     &     &     & \OK
  \\ \hline
  EmailSys & E2
     & \parbox[t][][t]{\hsize}{E-mail template \newline management}
     & .NET, MSSQL
     & audit
     & strict
     & \OK &     & \OK
  \\ \hline
  LendingSys & E3
     & Micro-lending
     & Ruby
     & flexibility
     & mutable
     &     & \OK & \OK
  \\ \hline
  ObjectSys & E4, E5
     & Object registration
     & JVM, Oracle
     & \parbox[t][][t]{\hsize}{audit, \newline flexibility}
     & strict
     & \OK &     & \OK
  \\ \hline
  VideoSys & E6
     & Streaming video
     & \parbox[t][][t]{\hsize}{JVM, EventStore, \newline Neo4J}
     & flexibility
     & mutable
     & \OK & \OK & \OK
  \\ \hline
  CMSys & E7
     & Content management
     & \parbox[t][][t]{\hsize}{PHP, CouchDB, \newline PostgreSql}
     & complexity
     & mutable 
     & \OK &     & \OK
  \\ \hline
 PaymentSys & E9, E10
     & Payment processing
     & \parbox[t][][t]{\hsize}{JVM, Groovy, \newline MongoDB, MySQL}
     & trending
     & mutable
     &     &     & \OK
  \\ \hline
 ApproveSys & E13
     & Approval processing
     & .NET, RavenDB
     & complexity
     & mutable
     & \OK &     & \OK
  \\ \hline
 MeetSys & E15
     & \parbox[t][][t]{\hsize}{Appointment\newline  management}
     & .NET
     & flexibility, complexity
     & mutable
     & \OK &     & \OK
  \\ \hline
 ProjectSys & E17
     & Project administration
     & \parbox[t][][t]{\hsize}{.NET, RavenDB, \newline PostgreSql}
     & \parbox[t][][t]{\hsize}{audit, \newline flexibility}
     & \parbox[t][][t]{\hsize}{cut-off \newline moments}
     & \OK &     & \OK
  \\ \hline
 IdentitySys & E20, E21
     & Identity management
     & PHP, MariaDB
     & \parbox[t][][t]{\hsize}{audit, \newline flexibility}
     & strict
     & \OK &     & \OK
  \\ \hline
 P-PaySys & E25
     & Payment platform
     & Golang, PostgreSql
     & \parbox[t][][t]{\hsize}{trending, \newline flexibility}
     & strict
     & \OK & \OK & \OK
  \\ \hline
 DocumentSys & E19
     & Document automation
     & .NET, MongoDB
     & \parbox[t][][t]{\hsize}{audit, \newline flexibility}
     & \parbox[t][][t]{\hsize}{cut-off \newline moments}
     & \OK & \OK & \OK
  \\ \hline
 Advert1Sys & E8
     & Classified advertising
     & JVM, MongoDB
     & \parbox[t][][t]{\hsize}{audit, \newline flexibility}
     & mutable
     & \OK &     & \OK
  \\ \hline
 Advert2Sys & E12
     & Classified advertising
     & .NET, MSSQL
     & trending
     & strict
     & \OK & \OK & \OK
  \\ \hline
 InventorySys & E11
     & Inventory management
     & .NET, LMDB
     & flexibility, complexity
     & mutable
     & \OK & \OK & \OK
\end{tabular}
\end{table*}

Event sourcing is applied in enterprise applications, either business-to-business or business-to-consumer, as illustrated by the interviews.
We did not encounter systems using event sourcing for IoT systems, or other stream processing systems.
This is in accordance with the community from which event sourcing originated, which focuses on enterprise applications.

The systems overview shows that the event sourcing pattern is not tied to a particular technology stack.
This diversity in technology confirms that event sourcing is indeed a pattern, and not a technology.

\subsection{Rationale for ESSs}

The reasons for applying event sourcing can be grouped into four categories.
Remarkably, all systems under study benefit from event sourcing, and no system returned to a current state model.
Still, most engineers state that they would not apply event sourcing in every system.
The reason given for this opinion is the added complexity of introducing event sourcing.
Engineer E2 would apply event sourcing by default, because of the benefits it gives.
 The different rationales as discussed with the engineers are summarized in Table~\ref{tab:rationales}.

\begin{table}
  
  \renewcommand{\arraystretch}{1.25}
  \caption{
    \small{The rationales given by the engineers, categorized in four concepts: audit, complexity, flexibility, and trend.}
  }
  \label{tab:rationales}
  \centering
  \footnotesize
  \begin{tabular}{p{1.3cm}|p{6.2cm}} 
     \textbf{Concepts}
     & \textbf{Codes}\\
  \hline
	Audit
	& Regulations (E4, E5, E14, E17, E19, E20, E21); \newline
	  Customer service support (E2, E4, E5, E7, E8, E9, E10, E12, E17, E18, E22, E23);  \newline
	  Explanation (E14, E23, E24)
	\\
	Complexity
	& Decoupling (E2, E11, E13, E16);  \newline
	  Distribution (E1, E3, E6, E12, E13);  \newline
	  Temporal logic (E2, E3, E13, E15, E16, E18, E24); \newline 
	  Process versus data (E4, E5, E7, E8, E9, E10)
	\\
	Flexibility
	& Multiple views on data (E3, E6, E7, E8, E14, E15, E17, E19, E20, E21, E23, E25);  \newline
	  Data is not discarded (E11, E24);  \newline
	  Data replication (E1, E4, E5, E6, E24);  \newline
	  Scalability (E2, E4, E5, E11)
	\\
	Trend
	& Experiment (E2, E12, E14, E25);  \newline
	  Learn (E1, E9, E10, E25)
  \end{tabular}
  
\end{table}

One of the main benefits of applying event sourcing is the retention of all state changes.
According to E24, event sourcing prevents prematurely data deletion: \engineer{as a software developer building a data-driven system and you are modifying data, you are essentially destroying your older copy of the data. And who told you you're allowed to delete data?}
We classified this group of rationale with the category \textbf{audit}~\citep{VanDerAalst2010} (9/19 systems).
Compliance with regulations (such as system ProjectSys) is one of the reasons in this category.
Improving customer support (ProjectSys, Advert1Sys) is another reason.
In those systems the state changes are used to explain the system and its behavior to customers.
Finally, simply explaining why and by whom data is changed (in debug scenarios for instance) is given as reason too (EmailSys).

The second category is \textbf{flexibility}~\citep{Lassing1999} (12/19 systems).
These systems chose event sourcing (and CQRS), because of the flexibility it provides in the architecture of the system.

Examples of this flexibility are the creation of secondary indexes for search (VideoSys), building and refreshing caches (B2CSys), replacing event queues (MarketingSys, WebBuildSys, LendingSys), and scaling out to multiple read databases (VideoSys).
Section~\ref{sec:eventstoreandsystem} explains how this flexibility is achieved through the implementation of different \emph{projections} and \emph{projectors}.

The third category is \textbf{complexity}~\citep{Biemans2001} (4/19 systems).
These applications were considered to contain complex business logic, heavily process driven instead of data driven.
Therefore, the architects designed the system as an event-driven system, starting out with the modelling of processes instead of data.

The final category, and only the rationale for three of the 19 systems, is \textbf{trending}~\citep{Clements1997} (3/19 systems).
The systems PaymentSys, P-PaySys and Advert2Sys started with event sourcing, because the (lead) architects picked up on a trend.
They were curious to the details of the pattern, and started to implement it in the new system.
In hindsight, the systems did benefit from this decision, although E9, E10, and E12 ascribe this to luck, and not to the design practices.

\subsection{Characteristics of Event Sourced Systems}

The core category of the GT process is \textit{the process of designing and implementing event sourced systems, as performed by software engineers}. 
As we needed to make sure that event sourced systems are not a technology but a technology agnostic pattern, we wanted to assure the types of applications and the technology platforms used to realize the implemented systems.
Three dimensions, the size of the event store, the workload handled by the application, and the size of the schema, are listed to indicate what kind of systems benefit from event sourcing.
These dimensions assure that event sourcing is not biased towards systems of a certain size.
Three related topics emerged during the coding process: DDD as software design approach, CQRS being a related architecture pattern, and the Microservice Architecture (MSA) style.
Together with the degree of immutability and the type of application, these different aspects from the characteristics that are listed in Table~\ref{tab:systems}.

Event sourcing is a pattern that stores every state change, immutability is thus at the core of the pattern.
\cite{Helland2015} states that immutability of data is a crucial aspect for distributed systems.
Although often seen as the defining characteristic of event sourcing, immutability is not enforced in any manner, as opposed to a blockchain.
In a number of the systems under study, immutability is sacrificed for a simpler schema evolution technique (see Section~\ref{sec:evolution}).
We observed different degrees of immutability.
The first degree is \textbf{strict}, 8 out of the 19 ESSs never change an event.
The second degree of immutability is used by 3 out of 19 systems, which allow for~\textbf{cut-off moments}.
In such a cut-off moment, the event store is changed, but back-ups guarantee that no information is deleted.
The goal of these back-ups is to satisfy regulations or service-level agreements, therefore, they are kept around forever.
This degree of immutability still guarantees an audit trail, because the back-ups can be used to retrieve all the state changes.
The last degree level of immutability is \textbf{mutable}, 8 out of 19 systems allow events to change.
In these systems, the event store is changed on some occasions, and the back-ups are not kept forever.
These systems do not satisfy the goal of a complete audit trail.
However, the events can still be used to explain how the current state was reached.
None of the ESSs lose information regarding the current state of a system.
Events that are changed, or transformed, are in most cases changed because of technical reasons.

In 14 of the 19 ESSs under study DDD is used as the design approach.
DDD is an approach to software development that aims at \emph{tackling complexity in the heart of software} (as the subtitle of the seminal book by~\cite{Evans2003} states).
DDD focuses on the explicit modeling of the domain, including its boundaries and events.
However, only four of the 25 engineers argue that DDD is a prerequisite for event sourcing.
Although the other engineers do not see DDD as a prerequisite, without a doubt DDD inspired the design of many ESSs.
Events, as expressed by E11, \engineer{should represent real world business events}.
This is different from transactional processing, or stream processing.
In those systems events can have a more technical nature.
An ESS that contains events not representing real world business domain events will undergo more changes to the software according to E11.
E11 explains: \engineer{You align the events with real world events, so you are dealing with changes that have a native equivalence. Doing DDD leads to a less fragile design.}
For E16, the understanding of the domain is a prerequisite for doing event sourcing: \engineer{A high level of maturity of the domain knowledge is a prerequisite. When the domain knowledge is still evolving, applying event sourcing introduces more risk.}

CQRS is a closely related pattern that also originated from the community around the DDD approach (the pattern itself will be explained in more detail in Section~\ref{sec:eventstoreandsystem}).
Although engineer E14 has seen a few solutions that apply CQRS without event sourcing, they are almost always used together.
All of the systems that we discussed with the engineers applied both CQRS and event sourcing.
The interviews give no explanation for this co-appearance.
A possible explanation, based on the experience of the authors, could be the fact that they are often `advertised' together in the community. 

Also closely related to event sourcing is the MSA~\citep{Dragoni2017} style.
Similar to DDD, the MSA style also attacks the complexity of large software systems.
This is confirmed by 8 of the 19 systems that were discussed in the interviews.
They implement microservices to break up a large application and control complexity by spreading the business logic over these services.
We observed two approaches in the systems that combine MSA and event sourcing.
The first approach uses event sourcing as an implementation detail of the microservices.
In the second approach, the events are not only used to store state changes, the event store is also used to communicate these events between microservices.

Unfortunately, the experts could not uniformly report on event store size, traffic, and schema size of the characterized ESSs.
Some of them could not disclose these details due to commercial reasons, while others no longer had access to the discussed system.
Table~\ref{tab:systemsizes} summarizes the details that were reported per discussed system.
The systems have a size ranging from smaller than three gigabytes, up to 250 gigabytes (or more than a billion events).
Eleven systems (including HealthSys that reports a growth rate of 4 million events per day) have more than a million events in the store, representing more than half of the systems.
Two systems (WebBuildSys and InventorySys) even report sizes over a billion events.
Advert2Sys shows a small event store size, but that is due to the active pruning that they do.
The growth of 4 million events per day shows that the total number of events is much higher than the reported five million.
The growth rate of the systems shows that a number of systems report a growth that passes the million events per day (HealthSys, Advert2Sys, and InventorySys), but most show a number far less than a million new events per day.
The schema sizes shows that none of the reported systems passes the 500 event types, but is rather spread out between 20 and 450 types.
Overall, Table~\ref{tab:systemsizes} shows a wide variety of event store sizes, handled traffic, and event store schema sizes.
Systems VideoSys, PaymentSys, ApproveSys, and InventorySys show that ESSs are not only used for small business domains.
And the event store size shows that event sourcing can be used for both small and large systems. 

\begin{table*}
  
  \renewcommand{\arraystretch}{1.25}
  \caption{
    \small{The size and growth of the event store and the schema size of the ESSs under study, as reported by the experts. Empty cells represent unknown data points.}
  }
  \label{tab:systemsizes}
  \centering
  \footnotesize
  \begin{tabular}{p{1.8cm}|p{7cm}|p{4cm}|p{2.5cm}} 
     \textbf{System Code}
     & \textbf{Event store size}
     & \textbf{Growth of the store}
     & \textbf{Schema size}\\
  \hline
  MarketingSys
     & $\geqslant$ 50,000,000 events 
     & 10,000 events per day
     &
  \\ \hline
  HealthSys 
     &
     & 4,000,000 events per day
     &
  \\ \hline
  WebBuildSys 
     & More than 100,000,000 active sites,
       every site owns hundreds or maybe even thousands of events
     & 
     & A single event stream type per site
  \\ \hline
  B2CSys 
     & $\leqslant$ 5 gigabyte
     &
     & 50 event types
  \\ \hline
  EmailSys 
     & \engineer{It is probably approaching the half a million events mark by now}
     & $\leqslant$ 50 or 60 events per day
     &
  \\ \hline
  LendingSys 
     & \engineer{We processed I think half a million account transactions}
     &
     & 6 microservices
  \\ \hline
  ObjectSys 
     & 200,000,000 events
     &
     & 50 event types
  \\ \hline
  VideoSys 
     & 7,000,000 events
     &
     & 400 event types
  \\ \hline
  CMSys 
     & $\leqslant$ 3 gigabyte
     &
  \\ \hline
 PaymentSys
     & 5,000,000 events
     &
     & 300 event types
  \\ \hline
 ApproveSys 
     & 1,000,000 events
     & 100,000 events per 2 months
     & 300 event types
  \\ \hline
 MeetSys 
     & 100,000 events
     & 1,000 events per day
     & 20 event types
  \\ \hline
 ProjectSys
     &
     & 10 events per minute
     &
  \\ \hline
 IdentitySys 
     & 50,000 events
     &
     & 20 - 30 event types
  \\ \hline
 P-PaySys 
     & \engineer{I don't think our scale is particularly high}
     &
     & 20 - 30 stream types
  \\ \hline
 DocumentSys 
     & 5,000,000 events
     & 1,000,000 events per month.
     & 100 event types
  \\ \hline
 Advert1Sys 
     & 50,000,000 events
     & 60,000 events per day
     & 115 event types.
  \\ \hline
 Advert2Sys
     & 5,000,000 events (active event store)
     & 1,000,000 events per day
     &  50 event types
  \\ \hline
 InventorySys 
     & 1,100,000,000 (250 gigabyte)
     & 77,000,000 events per month
     & 450 event types
\end{tabular}

\end{table*}

\section{Event Stores and Event Sourced Systems}
\label{sec:eventstoreandsystem}

\newtheorem*{event}{Event}
\newtheorem*{sequence}{Event Sequence}
\newtheorem*{stream}{Event Stream}
\newtheorem*{store}{Event Store}
\newtheorem*{eventschema}{Event Schema}
\newtheorem*{streamschema}{Event Stream Schema}
\newtheorem*{storeschema}{Event Store Schema}
\newtheorem*{conformse}{Conforms(E, $E_{sch}$)}
\newtheorem*{conformss}{Conforms(S, $S_{sch}$)}
\newtheorem*{conformses}{Conforms(ES, $ES_{sch}$)}
\newtheorem*{projection}{Projections}
\newtheorem*{project}{Project function}
\newtheorem*{accept}{Accept function}
\newtheorem*{projector}{Projector}

This section defines key concepts and operations in an event sourced system (ESS).
These definitions are based on our experience building ESSs, and confirmed by the interviews that were conducted.
They are used to conceptualize event sourcing and the identified challenges. 
When coding the interviews, different characteristics and variability of the concepts and operations were identified, which are described in this section.
These concepts and operations should be used in discussing, and teaching ESSs.

\subsection{The Event Store}

We propose the following definitions for the concepts and operations related to an event store.
First the concepts are defined, starting with events all the way up to the store.
After that the operations on the event store are given.

\begin{event}
  An event is a discrete data object specified in domain terms that represents a state change in an ESS.
\end{event}

An example of an event from the Netflix case~\citep{Avery2017} that represents a real world business event is given in JSON format:
{\small 
\begin{verbatim}
{ "LicenseCreated": { 
    "customerId": "BlackMirror", 
    "titleId": "TheNationalAnthemS01E01", 
    "date": "2014-01-06" } } 
\end{verbatim}
}
The importance of the relation to the business domain is stated by E5: \engineer{business analysts are telling us what the events should be.}
E11 adds \engineer{you capture business changes as a flow of events, you align these events with real world events.}
A more general definition is given by~\cite{Michelson2006}: ``a notable thing that happens''.
It lacks the relation to the business domain as it is used for event-driven architectures in general.
The data in the events can be stored in different formats such as JSON, XML, AVRO~\citep{TheApacheSoftwareFoundation}, or Protobuf~\citep{GoogleInc.}.
Events are stored in a sequence, in event streams.

 Both E14 and E25 do see a distinction between \emph{internal} and \emph{external} events.
Internal events are fine-grained and contain more detail, while external events are more coarse grained and meant for other systems to communicate.
Through this distinction it is possible to hide internal business logic from external consumers.
Multiple engineers (E12, E14, E16, E17, and E22) also acknowledge the usefulness of state propagation through events.
Instead of events that mark a business event, events can also be used to simply propagate the state of an object. 

\begin{sequence}
  Every event is stored together with a sequence number. Its sequence number represents the position of the event in the stream.
\end{sequence}

\begin{stream}
  An event stream $\textbf{s}$ is a sequence of tuples, each tuple containing an event and its sequence number \\ 
  \[s = \langle (\textnormal{e}_1, 1), (\textnormal{e}_2, 2), ..., (\textnormal{e}_n, n) \rangle\]
  The sequence numbers are consecutive natural numbers, starting with the number 1.
\end{stream}

The sequence numbers are not handed out by the event stream, but are supplied by the producer of the new events.
The event stream does validate if the sequence numbers are consecutive natural numbers.
E3 explains how this is used by event subscribers: \engineer{you get this monotonically increasing sequence of events that you can use to record your position.}
The streams together are stored in the event store.

\begin{store}
  An event store is a set of event streams. 
  These streams form the partitions of the event store, and are disjoint.
\end{store}

The event store has two foundational operations on event streams: $read$ and $append$.
The $read$ operation enables systems to read an event stream from a given sequence number.
Events are appended to the event stream with the operation $append$.
E20 explains how append is the only operation that changes the event stream: \engineer{I only append new events, and never throw away old events.}
The $append$ operation has an extra validation: the caller should supply the sequence number for the new event, which is validated and an error is returned if it is not the expected number.
Through this validation the store achieves \emph{optimistic concurrency control}.
According to engineer E24, this is the strongest guarantee that the event store should offer.
A caller will first need to $read$ from the event stream, before $append$ can be called.
If another caller calls $append$ in between, the $append$ of the first caller will fail, because the highest sequence number has changed.

Both the $read$ and $append$ operation operate on single streams, this emphasizes the fact that the streams in an event store are disjoint.
The $append$ function can either append a single event, or multiple events, depending on the implementation.
For instance \cite{EventStore2018} implements the $append$ function with a version that atomically appends multiple events to the stream.

\subsection{The Event Sourced System}

Enterprise software applications support at least two foundational use cases: storing information and retrieving information.
The event store is used to store the state changes in the system, however, the event store is not optimized for retrieving information.

In ESSs the $project$ function is central in both storing new information and retrieving of information.
First we define and characterize this $project$ function.
Second we discuss storing and retrieving information by presenting two parameterized operations.

\begin{project}
  The project function takes one or more event streams and creates a projection with the data from the given events. 
  The projection itself can take different forms, for instance it can be a relational database is updated through SQL statements, or a search index manipulated through the filesystem.
\end{project}

The $project$ function operates on one or more event streams.
The event streams are disjoint, and the $project$ function thus can not assume an order between the events from the different streams.
While the order of events in a single stream is guaranteed, the events from different streams have no relation.

The projection that is built by the $project$ function in an ESS is similar to the concept of projections in relational algebra~\citep{Date2003}: projections contain a selection of the data present in events.
Projections are similar to \emph{views} in a relational database: a selection and transformation of one or more database tables.

\begin{projection}
  A projection $\pi$ is a selection of the data stored in events, transformed into a specific model. 
  The selection and transformation depends on the purpose of the projection.
  The data in a projection is transient, a projection can be rebuilt from its source events at any point.
\end{projection}

Examples of different variations of projections are frequently given by the engineers.
Engineer E6 for instance explains how they project the event data to both Neo4J (a graph database) and ElasticSearch (a document database).
The graph database serves the navigation through the data, while the document database serves the search functionality.
Other examples given are a specific storage technology for indexing (used by for instance by E8, E12, E23), an analysis to report abuse of accounts information, and a relational table with all issued licenses for downloaded content.

The primary design question of the $project$ function and its target $projection$ is its purpose.
The importance of the $project$ function lays in its encapsulation of the variability in storage technology, data selection, and data model.
Choices can be made per project function, which enables a huge potential for optimized projections for their purpose.
The flexibility as reason for choosing event sourcing (Section~\ref{sec:challenges}) is in large part caused by the $project$ function.

The $project$ function also poses a risk for the performance of the system, a challenge we discussed in Section~\ref{sec:challenges}.
The time it takes to build a projection depends on two factors: the number of events that are read and the time it takes to update the projection.
Engineers E11, E13, and E14 discuss their search for improved implementations of projectors.
Quick improvements can be found in faster storage technology, or better use of hardware.
Engineer E12 explains how they prune the event stream by moving older events into a different stream.
This pruning decreases the number of events that the $project$ function needs to process, making the rebuilding faster.
Engineer E14 discusses how they rather plan the rebuilding in weekends, instead of investing developer effort for optimization.

The retrieval of information from the event store is done by building a projection.
Queries are answered using the data available in the projection.
Projectors can build the projection on-demand, or opportunistic: the given projection is build first and then the specific query is answered.
However, it is also possible to pre-build the projection: the projector constantly watches the event streams and updates the projection whenever new events arrive.
This decision depends on the ratio between reads of the projection and new events being appended to the stream.
If a projection is read infrequently, it is unnecessary to constantly project new events, and thus consume resources.
However, if a projection is read frequently projecting the new event directly on arrival improves the performance of the query.

The behavior of the projector is similar to that of the higher-order function \emph{fold}~\citep{Hutton1999}, a recursion operator that works on lists, as stated by~\cite{Meissner2018b}.
The projector \emph{folds} over the specific event streams and creates a projection.
The integration of functional programming and domain-driven design is further explored by~\cite{Wlaschin2018}.

Storing new information is done using the $append$ operation.
The $append$ operation is the only operation that is capable of storing new events in the store.
However, before storing these new events, they have to be produced.
Events in an ESS are produced as a result of an action (the commonly used name is \emph{command}) that is accepted by the system.
The validation, resulting in an acceptance or rejection of the command is done by the $accept$ function.

\begin{accept}
The accept function takes a projection $\pi$ and an command $c$. 
The command is validated using the data in the projection, and the accept function either results in an error or in an event.
\end{accept}

The command follows the \emph{Command pattern} described by~\cite{Gamma1995}.
The system first builds a projection, and then validates the command using the $accept$ function.
Validation of the command can result in either an new event or an error (in case of a validation error).
The new event is appended to a specific event stream, which is selected based on properties present in the command.
This appended event is the new information stored in the system.
While the projection is built in order to validate the command, it is only used to validate the command and is volatile.

A command can only affect a single stream, because the $append$ operation appends to a single stream.
To guarantee the consistency of information, the system should not append events to two streams in one request.
One append might fail, leaving the system in an inconsistent state.
This rule increases the importance of the design of the schema of an event store.

\subsection{The Schema}

An event store contains no schema for the specific structure of events.
The data schema is not explicitly defined at all, but is implicitly encoded in the ESS.
The knowledge of the data schema inside an ESS is encoded in the source code of the $accept$, and $project$ functions.
This is similar to other systems with a so-called implicit schema~\citep{Fowler2013}, such as document oriented data storage systems.

In general, events can take any form and thus the schema as well, therefore, we left these definitions abstract on purpose.
However, we believe that these abstract definitions can be used to support the discussion of schema evolution, as we show in Section~\ref{sec:evolution}.
This section defines event, event stream, and event store schemas, along with the \emph{conforms} relation.

\begin{eventschema}
  An event schema $\varepsilon$ describes the type and form of events.
  $\textit{conforms}(e, \varepsilon)$ holds if event $e$ conforms to the specification $\varepsilon$.
\end{eventschema}

An event schema could be implemented by for instance XML Schemas, or AVRO~\citep{TheApacheSoftwareFoundation}.
The latter uses the schema not only for validation, but also for serialization to a binary format.
Two other options that can be applied to create a more formal event schema are domain specific languages (suggested by E11 and E14) and strongly typed classes (see Table~\ref{tab:implementation}). 

\begin{streamschema}
  An event stream schema $\varsigma$ describes an event stream and the events that can occur in the stream.
  The event stream schema contains the event schemas of the events that can occur in the stream, along with the patterns of occurrence.
  $\textit{conforms}(s, \varsigma)$ holds if event stream $s$ conforms to the specification $\varsigma$.
\end{streamschema}

An event stream schema contains both the specification of the events, and the specific patterns.
An example schema contains both the schema (or specification) of the `registered' event, and the fact that the `registered' event occurs before a `checkout' event.

\begin{storeschema}
  An event store schema $\theta$ describes an event store and the streams that are stored in the event store.
  $\textit{conforms}(es, \theta)$ holds if event store $es$ conforms to the specification $\theta$.
\end{storeschema}

The event store schema contains more knowledge than only the event stream schemas, similar to the event stream schema. 
For instance the cohesion between streams, such as the fact that when a specific stream contains a certain event another stream should exist is also present in the event schema, can also be specified in the event store schema.
An explicit implementation of event stream schemas or event store schemas was not encountered during the interviews.

\subsection{Event Sourced Systems based on CQRS}

\begin{figure*} 
  \centering
  \includegraphics[page=1, clip, trim=25cm 7.5cm 27cm 0.5cm, width=\textwidth]{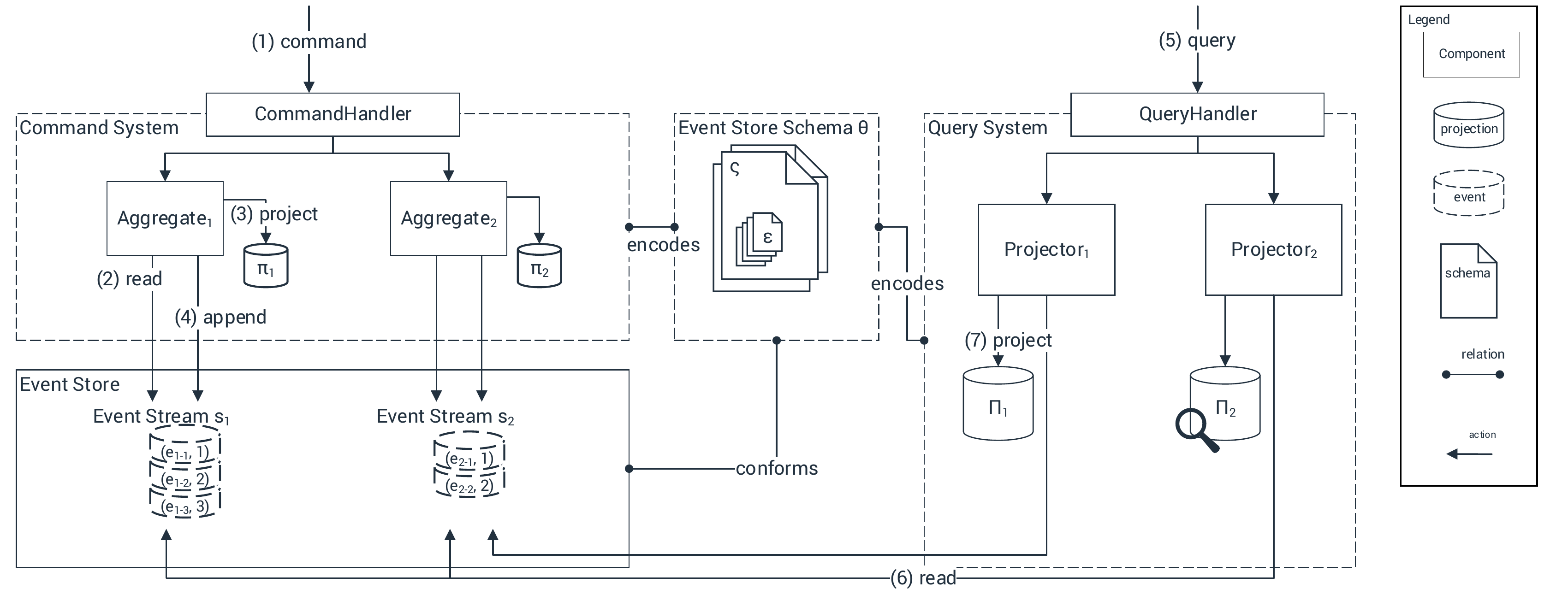}
  \caption{An event sourced system based on CQRS. The event store conforms to the schema $\theta$, which is encoded by the command and query systems. The command system validates commands using the events. These same events are read by the query system to build the projections, which are used to respond to queries. }
  \label{fig:event-sourced-cqrs-system}
\end{figure*}

As we have seen in Section~\ref{sec:esinpractice}, every ESS under study also applies CQRS.
CQRS was introduced by \cite{Young2010} and \cite{Dahan2009}, and the goal of this pattern is to separate actions that change data (those are called commands) from requests that ask for data (called queries).
Although event sourcing and CQRS can be used separately, the common application of the two patterns is worth exploring.
Based on literature and the interviews an example architecture combining event sourcing with CQRS is discussed.
This architecture is shown in Figure~\ref{fig:event-sourced-cqrs-system}.
As illustrated, the event store schema $\theta$ is part of the ESS: the event store conforms to it, and the command and query system encode it in their application logic.

In the command system $aggregates$ (as introduced by~\cite{Evans2003}) are used to process incoming commands (1).
Commands are routed by the $commandhandler$ to the correct aggregate.
Aggregates will process the commands using the $accept$ and $append$ operations.
First the existing events are read (2), a projection is built (3), and then the $accept$ function will be called.
When the command is accepted, the resulting event will be appended to the event stream (4).

An aggregate reads a specific event stream, to which the new event is also appended.
Often the aggregate will be the owner of the event stream it reads and appends to.
As a benefit, commands sent to different aggregates can be processed concurrently without interfering.
E6 describes a solution were multiple aggregates use the same stream.
This variation is used to share generic behaviour among aggregates, it is mixed with more specific logic.

In the query system, $projectors$ are used to build $projections$ that can be used to return information to the sender.
Queries are routed by the $queryhandler$ to the correct projector (5), depending on the specific purpose of the projector (such as browsing or searching).
The projector will retrieve the requested information from its projection.
First the events from the event streams will be read (6), then the projection will be built (7).

While queries can be handled by building the projection on-demand, most ESSs based on CQRS will update the projection as soon as new events are appended.
In that scenario, step (6) and (7) will be executed before (5), and the projector can immediately use the projection to handle the query.
This decision is based on the ratio between events and queries.
When there are few queries, and many events, pre-building the projection takes up resources (such as storage).
If the workload consists of more queries, building the projection ahead of time results in faster response times.
 E24 describes a flexible approach that merges the two approaches in an on-demand fashion. 
The sequence numbers of events are used as checkpoints and allow the projectors to track which events are already processed.
Immutability of the event store is crucial for these projectors, if events or their ordering are mutated, the checkpoint has no value and the projector needs to re-read the event streams and rebuilt the projection.

Most pre-built projectors are \emph{eventually consistent}.
As \cite{Vogels2009} explains, the ESS guarantees that if no new commands are processed, eventually all queries will return the last updated value.
However, because there is time between the acceptance of a command and the updating of a projection, a query might return an older value.
The duration between (4) and (7) is the so-called inconsistency window: the command system and the query system do not share a consistent state.
Eventual consistency was also listed as one of the challenges in ESSs and is discussed in Section~\ref{sec:challenges}.

Four engineers explain how there projectors share a database transaction with the aggregates.
This allows them to achieve immediate consistency, because both the event as the projections are committed in a single projection.
In those systems scalability is sacrificed for immediate consistency.
This implementation technique results in \emph{synchronous} projections.

 Table~\ref{tab:implementation} summarizes the different concepts and codes that were extracted from the interviews.
While the definitions are mainly based on our experience in building an ESS, we have used the data extracted from the interviews to scope our description.
The concepts and codes discussed by the engineers determined what specifics were described. 

\begin{table}
  
  \renewcommand{\arraystretch}{1.25}
  \caption{
    \small{The concepts and codes extracted from the interviews 
           related to the implementation of CQRS based ESSs.}
  }
  \label{tab:implementation}
  \centering
  \footnotesize
  \begin{tabular}{p{2.7cm}|p{4.8cm}} 
     \textbf{Concepts}
     & \textbf{Codes}\\
  \hline 
  Event Store &
      Business events (E5, E11); \newline
      State propagation (E12, E14, E16, E17, E22); \newline
      Monotonically increasing sequence number (E3); \newline
      Append only (E1, E2, E16, E17, E20); \newline
      Optimistic concurrency control (E24); \newline
      Internal versus external (E14, E25)
    \\
    Event Sourced System &
      Projector variations (E6, E8, E12, E23); \newline
      Optimization of projecting (E11, E12, E13, E14)
    \\
    Schema &
      Domain Specific Languages (E11, E14); \newline
      Strongly typed classes (E2, E3, E4, E5, E17)
    \\
    CQRS: Projections &
	  Synchronous (E2, E20, E21, E23); \newline 
	  Opportunistic (E24);  \newline
	  Independent (E16, E17)
	\\
	CQRS: Aggregates &
	  Multiple on one stream (E6);  \newline
	  Snapshots (E2, E20, E21); \newline
	  Instance versus type (E14, E25)
  \end{tabular}
  
\end{table}

\section{Challenges Faced in Applying Event Sourcing}
\label{sec:challenges}

A pattern description without discussing the consequences is incomplete, and would lead engineers astray.
While Section~\ref{sec:esinpractice} discusses the positive consequences that engineers experienced, they also discussed the negatives in the interviews.
In this Section we discus five challenges experienced by the engineers with two goals in mind: (1) to indicate to practitioners what the limitations of the pattern are and (2) to formulate novel research topics for future research around the pattern.
The first two challenges are addressed in more detail by two of our contributions in Section~\ref{sec:eventstoreandsystem} and \ref{sec:evolution}.
The summary of mentioned challenges by engineers is listed in Table~\ref{tab:challenges}.

\begin{table*}
  
  \renewcommand{\arraystretch}{1.25}
  \caption{
    \small{The challenges faced by the practitioners while implementing ESSs.}
  }
  \label{tab:challenges}
  \centering
  \footnotesize
  \begin{tabular}{p{6.5cm}|p{9.5cm}} 
     \textbf{Challenge}
     & \textbf{Codes}\\
  \hline
    How can Engineers better be Supported in Learning how to Apply the Event Sourcing Pattern?
	& Eventual Consistency (E1, E2, E14, E24);  \newline
	  Events versus state (E1, E2, E4, E5, E6, E12, E13, E15, E16, E17, E20, E21, E23, E25);  \newline
	  Lack of knowledge sharing (E1, E3, E9, E10);  \newline
	  Start is slow (E4, E5)
	\\
	How can Tools, Frameworks, and Platforms be Provided to Make the Pattern even More Successful?
	& Immature tools (E2, E9, E10, E17, E20, E21, E25);  \newline 
	  Frameworks not properly maintained (E6, E19, E22);   \newline
	  Pattern versus framework (E6, E24);   \newline
	  Tools not accepted by operations (E14);   \newline
	  Frameworks hide details from developers (E24);   \newline
	  Frameworks help beginners (E6, E7, E13, E14, E17, E24)
	\\
	How can Projections be Optimized?
	& Rebuilding is slow (E4, E5, E8, E9, E10, E13, E20, E21, E22, E25); \newline 
	  First in-memory (E8, E24);  \newline
	  Targeted rebuilds (E8, E9, E10, E16, E17, E20, E21, E23); \newline 
	  Rebuild versus developer time (E1, E2, E6, E7, E9, E10, E11, E13, E14, E15, E20, E21, E22)
	\\
	How can a System that Uses Event Sourcing Protect User Privacy?
	& Separate events from personal information (E20, E21); \newline 
	  Remove (E11, E23); \newline
	  Anonymization (E11, E25)
	\\
	How can Event Stores be Evolved?
	& See Table~\ref{tab:techniques-summary}
  \end{tabular}
  
\end{table*}

\noindent \textbf{How can Engineers better be Supported in Learning how to Apply the Event Sourcing Pattern? -}
The most prominent category of challenges mentioned by the engineers is in the area of designing software.
Designing ESSs is more difficult than other systems, because of two characteristics.
In the experience of thirteen of the 25 engineers, thinking in events and state transfers is completely different from thinking in current state and database transactions.
Section~\ref{sec:eventstoreandsystem} proposes a description that improves the understanding, and support the teaching of event sourcing and event sourced systems (ESSs).

However, an ESS introduces not only events and state transfers.
Eventual consistency forces developers to let go of guarantees that they would have in a system using current state and synchronous processing.
In a CQRS system, an update sent through a command will not immediately be reflected in the result of a query.
The system first needs to process the event into one or more projections.
Engineer E12 states that \engineer{a lot of developers had to get used to information not being in place}, and E2 adds that \engineer{getting people to understand eventual consistency is the biggest hurdle.}
Eventual consistency forces developers to rethink the basic interactions of the user with the system.

We give two examples of interactions that force developers to rethink system design.
The first example is that of the expectation of users to retrieve data that they previously submitted into the system.
However, in a CQRS system, the query system might not directly return the data that was submitted through a command.
The user interface of the system should make it clear to the user what is going on, or even try to hide the fact that the system is eventual consistent.
The second example is that of developers that more or less have the same expectation.
Often developers try to use the result of the query to make decisions in an aggregate.
However, the query system might not have processed all events and misses recent updates.
If developers overlook this principle, the decisions lead to bugs in the system. 

\noindent \textbf{How can Event Stores be Evolved? - }
Both E13, \engineer{we dreaded the upgrading, we had some fear in advance}, and E22, \engineer{versioning in event sourced systems is a big problem}, point out the perceived difficulty of upgrading ESSs.
This challenge did not come as a suprise, our earlier work~\citep{Overeem2017b} and the work of~\cite{Young2017} underline this.
During the interviews we identified five fundamental techniques for schema evolution in ESSs, which are described in Section~\ref{sec:evolution}.

\noindent \textbf{How can Tools, Frameworks, and Platforms be Provided to Make the Pattern even More Successful? - } 
Eight engineers discuss the lack of standardized tools, such as frameworks, platforms, and databases.
A commonly stated opinion within the community is that you do not need frameworks to implement an ESS.
However, engineers E9, E10, E17, E20, E21 and E25 state that they wish to see more mature libraries and frameworks.
Engineers E6, E17, E19, and E22 mention that infrastructure and tooling for ESSs is immature.
Either the tooling does not support a broad enough set of scenarios, or the quality is lacking.
How large the market is for specialized event sourcing tools is difficult to say.
Recently~\cite{AxonIQ2018} has started to offer commercial support for ESSs, similar to what~\cite{EventStore2018} does.

\noindent \textbf{How can Projections be Optimized? - }
Projections, as discussed in Section~\ref{sec:eventstoreandsystem}, are used to retrieve information from the system.
Rebuilding projections, however, can become a bottleneck for ESSs.

Engineers E11, E13, and E14 discuss their search for improved implementations of projectors.
Quick improvements can often be found in faster database technology, or better use of hardware.
Although rebuilding projections needs planning, engineer E14 discusses how they rather plan the rebuilding in weekends, instead of investing developer effort for optimization.

Engineer E16 explains how the domain can show an optimization: not reading all the events on a rebuild.
Often the older events are no longer reflected in the projection, because the specific data (such as a classified advertisement) is no longer active.

Another important implementation detail that lifts some of the burden is that projectors can (and must) be implemented as independent, autonomous processes.
This gives the system the possibility to only rebuilt those projections that are required to, instead of all the projections at once.

\noindent \textbf{How can a System that Uses Event Sourcing Protect User Privacy? - }
Privacy regulations, such as the GDPR, are designed to protect users from being taken advantage of.
Personal information should not be kept in a system for all eternity, but the system should delete it whenever someone requests that.
However, such a requirement conflicts with the nature of event sourcing: retaining all the data.
Engineers E20, E21, E23, E25 mention that they designed their systems to comply with these regulations.
Systems HealthSys and P-PaySys use some form of anonymization and removal of information to comply.
Obviously, this requires them to rewrite events.
System IdentitySys takes a completely different approach.
The system separates the events and the personal information in two different stores.
When events are read, they are supplemented with the personal information.
If that information is no longer present (because of removal requests), default values are supplied.

\section{Schema Evolution in Event Sourced Systems}
\label{sec:evolution}

A challenge discussed by multiple engineers is evolution of event sourced systems (ESSs) (as stated in Section~\ref{sec:challenges}).
From the transcripts, we identified five fundamental techniques for schema evolution.
These event schema evolution (ESE) techniques are described using the definitions given in Section~\ref{sec:eventstoreandsystem}.

We encountered two reasons why event schema evolution in ESSs is difficult.
First of all, the implicit schema (as described by~\cite{Fowler2013}) makes evolution in ESSs difficult.
Solutions as proposed by~\cite{Meurice2016} and~\cite{Maule2008} to analyze the impact of schema changes are not usable, because there is no explicit schema.
In contrast to their solution, the change originates in the application and impacts the data in the event store.
This makes the direction of the impact different than theirs.
 
The second difficulty in event schema evolution, is the immutability of the event store.
Traditional solutions to transform or rewrite the store are not always possible.
However, the benefits of immutability in event stores (as listed in Section~\ref{sec:esinpractice}) are not always requirements.
The different degrees of immutability, as shown in Table~\ref{tab:systems}, allows for different evolution techniques.

Teams that apply event sourcing without a clear understanding of the business domain introduce risk, according to E14, E16, and E22.
E22 explains that the challenge of evolution is exactly why it is preferred to always start a new system without event sourcing, and only introduce event sourcing when the domain knowledge is stable: \engineer{once we have enough trust in our model we will transform to event sourcing.}
As E16 confirms, events based on a sufficiently clear domain knowledge will decrease schema evolution.

Another prevention technique is the cleaning up of events in the event store, of which we encountered two possibilities.
First of all, older events that no longer represent active information can be moved into \emph{cold} storage.
These events can still be read and processed, but are no longer processed by the ESS itself.
Therefore, they do not have to conform to the implicit schema of the ESS.
Second, sometimes these events can be kept in the event store itself, but the ESS will never read them.
Again, this makes it possible to ignore those events on upgrades.

Event schema evolution that can not be prevented can be solved by the following five evolution techniques.
 Although in our work~\citep{Overeem2017b} we also discuss five techniques, during the interviews a different set of techniques was encountered. 
The technique \emph{lazy transformations} was not mentioned by any of the engineers, while \emph{weak schema} was mentioned as a new technique.
 Which techniques are used by which engineers, and the benefits and liabilities per technique given by the engineers during the interviews are classified in Table~\ref{tab:techniques-summary}.
In some cases the liabilities are also from engineers that do not apply the particular technique: they stated the liability as a reason for not using the technique. 

\begin{table*}
  \renewcommand{\arraystretch}{1.3}
  \caption{\small{Benefits and liabilities of event sourcing evolution techniques.}}
  \label{tab:techniques-summary}
  \centering
  \footnotesize
  \begin{tabular}{p{2.0cm}|p{2.7cm}|p{5.5cm}|p{5.5cm}}
                               \textbf{Technique}   
                             & \textbf{Engineers}   
                             & \textbf{Benefits}      
                             & \textbf{Liabilities}                                       
                             \\
     \hline
     Versioned Events        & \textbf{2}: E7, E19                   
                             & Simplicity of implementation (E19)        
                             & Application logic pollution (E7, E9, E16)
                             \\
     \hline                                 
     Weak Schema             & \textbf{11}: E2, E7, E8, E11, E14, E15, E16, E17, E20, E21, E22
                             & Simplicity of implementation (E2, E8, E11, E15, E17, E22)           
                             & Application logic pollution (E9) \newline
                               Feature incomplete (E8, E15, E17)                                       
                             \\
     \hline                               
     Upcasters               & \textbf{12}: E1, E4, E5, E7, E11, E12, E13, E14, E16, E19, E23, E24
                             & No application logic pollution (E19) \newline
                               Strict immutability (E24) \newline
                               Simplicity of implementation (E14)                               
                             & Decrease of run time performance (E11, E23) \newline
                               Multiple schemas (E23) \newline
                               Complexity of implementation (E23)                                       
                             \\
     \hline                            
     In-Place \newline Transformation
                             & \textbf{5}: E8, E9, E10, E13, E23
                             & Ad-hoc evolution (E8, E9, E10, E13, E23) \newline
                               Single schema (E13)
                             & Mutability of events (E22) \newline
                               Complexity of implementation (E13) \newline
                               Decrease of evolution performance (E24) \newline
                               Risk of data-loss (E8)                   
                             \\
     \hline                            
     Copy-Transform          & \textbf{14}: E3, E6, E7, E8, E9, E10, E11, E13, E14, E15, E17, E19, E22, E23
                             & Simplicity of implementation (E6, E13) \newline
                               Strict immutability (E15, E17, E19) \newline
                               Ad-hoc evolution (E3, E6, E17, E23)
                             & Mutability of events (E11, E16, E22) \newline
                               Decrease of evolution performance (E6, E24) 
  \end{tabular}
\end{table*}

\noindent \textbf{ESE Technique 1: Versioned Events -} Given an event store $es$ conforming to a schema $\theta$, the technique \emph{versioned events} transforms the schema into $\theta'$ such that
\[\textit{conforms}(es, \theta') \land \forall \varsigma \in \theta: \exists \varsigma' \in \theta': \varsigma \subseteq \varsigma' \]

This techniques introduces only new types of events, and does it in such a way that the event store $es$ conforms to $\theta'$ without transformation.
The $project$ functions that process the involved streams are required to handle these new events.

\findings This technique is applied by engineers E7 and E19, with the sole benefit that it is a simple technique that does not require specific changes to the ESS.
The liability of this technique is the pollution of application logic, as stated by E16: \engineer{I try to keep my domain abstraction pure. My v1 and v2 version of the event do not enter the model together.}

\noindent \textbf{ESE Technique 2: Weak Schema - } With this technique the events are described in a minimalistic manner. 
Similar to technique 1, the event store $es$ or the schema $\theta$ are not transformed during evolution.
Evolution operations that are allowed with this technique are limited to transforming the event $e$ into $e'$ such that it still conforms to the event schema $\epsilon$.
This requires the $project$ operation to handle this variability.

\related This technique is described by~\cite{Daigneau2011} as the \emph{tolerant reader} pattern.
Serialization formats such as Protobuf by~\citet{GoogleInc.} and AVRO by~\citet{TheApacheSoftwareFoundation} support this technique by reading the existing binary data into the new version of the objects.

\findings Eleven engineers apply this technique, because of the simplicity.
The limitations of this technique are stated as a liability, together with the pollution of the $project$ operation that is required (E9 explains: \engineer{you want to assume a certain event schema}).

\noindent \textbf{ESE Technique 3: Upcasting - } This technique is well known to event sourcing practitioners and described by~\cite{Betts2013}.
The event streams are transformed into streams conforming to the latest schema by a new function: the $upcast$ function.
This function is called before the streams are passed into existing $project$ functions.
The transformation is centralized in this new function, which improves the maintainability of the system.

For the $project$ functions it appears that little has changed, it appears that the relation $\textit{conforms}(es, \theta^{'})$ holds. 
However, events already stored in $es$ still $\textit{conforms}$ to $\theta$, while newly appended events conform to $\theta^{'}$.
After appending new events to $es$, the store itself will neither conform to $\theta$ or $\theta^{'}$.

\related The technique is similar to \emph{message translators} as described by~\cite{Hohpe2004}.

\findings Twelve engineers use upcasters, claiming benefits as no domain pollution, immutability of events, and simplicity of implementation.
One of the stated liabilities is an decrease in performance: \engineer{If you have been running upcasters for a long time, you will have quite a stack of them in place, which slows down the entire loading.}
Other liabilities are added complexity in analyzing the event store, because it contains events that conform to different schemas.

\noindent \textbf{ESE Technique 4: In-Place Transformation - } This technique updates events to resemble the new schema, and thus forces ESSs to forgo of immutability.
New operations that alter event streams need to be introduced, such as $insert$ (insert an event at a specific position) and $update$ (update the event at a specific position).
These operations break the immutability of the event store, with the consequence that cached projection need to be rebuilt.
Therefore, two available event stores, EventStore~\cite{EventStore2018} and AxonDB~\cite{AxonIQ2018}, deliberately do not offer these operations.

\related This technique is similar to migration scripts for relational databases.
\cite{Scherzinger2013} and \cite{Saur2016} both propose a similar approach to evolve data in a NoSQL store.
The lazy migration (on data access) is similar to \emph{incremental migration} as described by \cite{Sadalage2012}.

\findings Four systems, HealthSys, PaymentSys, ApproveSys, and Advert1Sys, apply this technique.
Benefits are the possibility of ad-hoc fixes, and improved reasoning because the store will only contain events conforming to a single schema.
However, the risk of making errors, the loss of immutability, and the performance are stated as liabilities.
E22 explicitly prevented this technique from being used: \engineer{to prevent this technique we first zipped the events, and then encoded the result before storing them.}

\noindent \textbf{ESE Technique 5: Copy-And-Transform - } During the execution of this technique, existing streams are processed and new streams are created from transformed events that conform to the new schema.
This does not violate the immutability of the source events, but creates new events instead.
Existing projections are still valid, although they do need to process new streams to receive new events.

\related \cite{Young2017} describes this technique as \emph{copy and replace}.
The \emph{parallel universe} of IMAGO, as described by~\cite{Dumitras2009}, is similar to this technique.
QuantumDB, created by~\cite{Jong2015b}, uses \emph{ghost} tables to apply this technique in relational databases.
Copy-and-transform of a complete event store could be seen as an ETL process that creates a new store.

\findings Fourteen engineers have used this technique, either to transform specific streams or a complete event store.
As E6 states, this technique is relatively simple to implement, because \engineer{we can do literally anything we want.}
The data preservation is stated as a benefit, as well as the fact that this is a one-time operation.
The performance of this operation is a liability, transforming a large store takes a considerable amount of time.

 The data discussed in Table~\ref{tab:techniques-summary} does not allow us to discuss how techniques are combined within a single system.
It does allow us to discuss how engineers have experienced and applied different techniques over the course of working on multiple systems.
We can observe the following from the discussed engineering experiences:
\begin{itemize}
    \item No engineer has solely applied \emph{versioned events} or \emph{in-place transformation}, those techniques are clearly used in combination with others.
    \item Five engineers have solely applied \emph{upcasters}, which corresponds with the general advice we found in the grey literature and community.
    \item The \emph{copy-transform} technique is mostly used in combination with other techniques, only two out of the fourteen engineers have solely applied this technique. 
    \item Four engineers have considered techniques, but opted to not apply them: E9 considered \emph{versioned events} and \emph{weak schema}, E16 considered \emph{versioned events} and \emph{copy-transform}, E22 considered \emph{in-place transformation}, and E24 considered \emph{copy-transform} and \emph{in-place transformation}.
\end{itemize}

We conclude that the techniques are not exclusive: almost all engineers have used multiple techniques and applied multiple techniques in a single system.
Example combinations mentioned in the interview are
\begin{itemize}
    \item The application of \emph{upcasters}, with \emph{copy-transform} to clean up the upcasters when there are to many.
    \item The application of \emph{in-place transformation} for quick patches, while a different technique is used for planned evolution.
    \item The application of \emph{weak schema} for simple evolution steps, while a different technique is used for more complex evolution.
\end{itemize} 

From the study we formulate the following advice:
\begin{enumerate}
\item \emph{Versioned events} and \emph{weak schema} are the simplest techniques to implement. Systems should start out with those techniques.
\item When evolution operations can not be handled by the first two techniques, systems can apply \emph{upcasting}. This retains the immutability of the event store.
\item Only when a decrease of performance or maintainability is experienced should systems apply \emph{copy-and-transform}.
\item \emph{In-place transformation} should only be used by those systems that do not require immutability or an audit log.
\end{enumerate}

The techniques form a range of possibilities to evolve the event store of an ESS.
All techniques with one exception, \emph{in-place transformation}, can be applied in an ESS that follows the definition given in Section~\ref{sec:eventstoreandsystem}.

\section{Discussion}
\label{sec:discussion}

One could wonder whether another research approach would have been equally successful in extracting architecture knowledge about the event sourcing pattern. 
We have looked at open source systems such as \citet{Axxon2016}, \citet{EventStore2018}, \citet{NEventStoreDevteam2018}, \citet{ProophComponents2018}, and observed that these follow the pattern and guidelines as discussed in this article. 
However, aspects such as the rationale and consequences of using the pattern are impossible to extract this way. 
This research is also similar to a study with multiple cases (\cite{Flyvbjerg2006}), although one would expect a more extensive extraction of information about the case (i.e., system) and its context in a multiple case study. 
We would have had to use more research resources, but perhaps we would have also been able to provide more code examples of how the pattern was implemented. 
Finally, design research (\citet{Sein2011}) could have also been used to extract the pattern description. 
While the description would perhaps have been less extensive, there would have been more focus on the evaluation and validation of the pattern and its description. 
We consider this last aspect as future work, even though we are convinced that the incremental nature of this research has led to a pattern description that is reusable and useful for architects.

Our pattern description itself does not follow a specific format.
We decided to structure our presentation according to the concepts emerged from the GT, and not according to a specific pattern description format.
We did, however, use the examples of~\citet{Gamma1995} to evaluate the completeness of our pattern description.

Gamma et al. states three essential elements besides the \textbf{the pattern name}: the \textbf{problem}, the \textbf{solution}, and the \textbf{consequences}.
The problem describes what the context is of the pattern, and when to apply it, which we have summarized in Section~\ref{sec:esinpractice}.
The description of the pattern, the solution, is covered in Section~\ref{sec:eventstoreandsystem}.
Finally, the consequences, are split over two sections: Section~\ref{sec:esinpractice} covers the positive consequences by linking them to the problems that are solved.
Section~\ref{sec:challenges} covers the negative consequences by stating several research challenges for future work.

The format that Gamma et al. use to describe patterns consists of thirteen different sections.
While these sections cover the four essential elements, the \emph{related pattern} section should be discussed on its own.
The design of a software system is never the application of a single pattern, but rather the combination of different patterns that together form the design.
This is not different in ESSs.
Section~\ref{sec:eventstoreandsystem} recognizes this, and explains the combination of event sourcing in CQRS in great detail.
The relation to other patterns to solve the specific challenges of schema evolution are covered in Section~\ref{sec:evolution}.

A second question that must be asked is whether academic fora are the optimal place to publish patterns. 
As whole books have been written about particular patterns and as patterns appear to have a certain shelf life, one could wonder whether patterns should be published in academia at all. 
We argue, with this article, that some patterns are too important to ignore (SOA, Client-Server, Event Sourcing, etc.) and that these deserve specific detailed attention from academics. 
 We find the strongest proof for this in the provided research challenges (Section~\ref{sec:challenges}) and in the challenge discussion about evolving event sourced systems (Section~\ref{sec:evolution}).

The number of interviews does not allow use to generalize over the results.
It is not possible to prove that, because 14 engineers use the technique \emph{weak schema} it is the recommended technique.
However, practitioners can integrate the reported experience into their decision making.
They can weight the context of the interviewed engineers, and match that with their own context.
Although our research does not result in hard recommendations, we believe that practitioners can benefit from the reported experiences. 

\section{Threats to Validity}
\label{sec:threats}

Both \cite{Golfasni2003} and \cite{Onwuegbuzie2007} discuss the challenges of assessing validity in qualitative research. 
We identify several biases for both internal and external validity. 
First, we regard the objects of study, i.e., the engineers and their uses of and experience with the pattern. 
The contributions of our research are based on the 25 interviews that were conducted.
The engineers were not hand selected, but volunteered.
Therefore, it is possible that we only interviewed a particular subset of practitioners, who are willing and able to discuss the pattern at length.
It is for instance remarkable that they all combine CQRS with event sourcing.
Table~\ref{tab:engineers} shows a diverse variety of experiences, and Table~\ref{tab:systems} shows an equally diverse variety of systems.
We have interviewed consultants (E14 and E16), and full-time employees, with a wide range of years of experience.
From small systems to multi-million user systems, the interviewed engineers have been exposed to all.
These characteristics indicate a broad range of opinions and experiences.
Within the group of 25 engineers, 16 engineers have three years or less of experience working on ESSs.
This could be due to the relative novelty of the pattern.
However, these engineers were full-time involved in the development of the ESS.
The exploratory questions~(Appendix~\ref{sec:interviewprotocol}) focus on topics that can sufficiently be answered by engineers with one or two years of experience.

Internal validity, which is strengthened by the way in which the research is conducted, has been defended in several ways. 
First, an interview and analysis protocol~(Appendix~\ref{sec:interviewprotocol}) has been applied to each interview. 
The interview protocol was created from extensive literature study and discussion in the research team, in which two members have no experience with the pattern itself, thereby reducing bias. 
The first two authors have extensive experience in developing a large ESS. 
This experience has lead to many interactions with practitioners in gatherings, conferences, and on-line.
These interactions have served as an informal triangulation that support the findings presented in this article.

As a constructivist GT approach~\citep{Charmaz1996} was followed, we conducted relatively open interviews. 
The exploratory nature of the interviews enabled interviewees to comment on all aspects of the subject under study, independent of the experience of the engineer with the pattern. 
Many engineers work on closed source, commercial systems, which makes it hard to use documentation or source code in the research.
Every interview was closed with the question if anything important was left unasked, and if they knew other engineers that we should interview.
Often the engineers came with stories and anecdotes that amplified the discussed topics.
The engineers that were referred us to were all invited to cooperate.

External validity, i.e. generalizability to other cases, can be defended by the multitudes of systems that the engineers have observed and worked on. 

As already discussed in Section~\ref{sec:research-approach} we do not claim to have reached saturation.
Not reaching saturation could leave us open to missing crucial information, or even using incorrect information.
Seven of the interviewed engineers have five or more years of experience, and we did not find conflicts between their statements and the other interviews.
Together with the experience of the first two authors in developing ESSs, we believe that our findings are supported by the data.

We have not covered all niches in the software world, so we can not generalize to all types of systems. 
However, we do believe that in the domain of business information systems, we have sufficient coverage to claim generalizability to other systems in this domain. 
Furthermore, while we do not claim generalizability to other domains, we do believe that those domains can be inspired by our findings in designing event sourced systems. 
Also, the common occurrence of all event sourcing evolution techniques in Table~\ref{tab:techniques-summary}, illustrates that we observed a broad cross section of systems in use. 
Finally, the use of GT has provided us with a reliable manner of extracting concepts and definitions from the interviews. 
While this study's findings can be generalized to describe event sourced patterns, the research work is not finished. 

\section{Conclusion}
\label{sec:conclusion}

In this article we present a conceptualization of the event sourcing pattern, grounded in interviews with 25 event sourcing engineers.
Event sourcing is a pattern that solves the three problems that modern systems face.
The flexibility that the combination of event sourcing and CQRS gives decreases the complexity in large systems.
The decrease of complexity enables the development of larger systems that remain maintainable.
The reliability of the system improves when every state change is stored in a durable store.
It allows engineers to undo state changes that were incorrect, or replay those state changes after system failures.
An improved reliability is essential for systems that provide increasingly critical processes.
Finally, systems that serve increasing numbers of end-users benefit of the improved scalability that ESSs systems provide.

These benefits give enough reason to incorporate event sourcing in modern systems.
This article presents a thorough description of the pattern, including the context in which it is applied and the consequences that are encountered.
The description itself is grounded in the experience of 25 engineers, making it a reliable source for both new practitioners and scientists.
We answer the following four research questions in this work.

\vspace{.3cm}
\noindent \textbf{What types of systems apply event sourcing, and why?} 
 The overview of 19 systems, given in Section~\ref{sec:esinpractice} and especially in  Tables~\ref{tab:systems} and \ref{tab:systemsizes}, show that event sourcing can be applied in systems of any size: both smaller and larger systems benefit from the pattern. 
We studied systems with thousands of events up to and including systems with billions of events, and all of these systems have benefited from event sourcing, according to their engineers.
As E14 states \engineer{I have never seen an event sourced system that was rewritten to a system with traditional current state storage.}
The event sourcing pattern is not tied to a specific type of application, but is applied in many different domains, such as marketing, micro-lending, content management and classified advertising.
The systems under study show a strong relation to DDD as a software development approach.
This is partially explained by the fact that event sourcing and CQRS were invented in the community that grew around DDD.
The microservice architectural style has a weaker relation (8 out of 19 systems apply it), while CQRS is used in all these systems.
We identify four reasons for event sourcing: audit, flexibility, complexity, and trending.
While a common characteristic of event sourcing is the immutability of the events, we show that there are three levels of immutability that can be found in ESSs.
The characteristics summarized in~\ref{tab:systems} substantiate that event sourcing can be applied in a diversity of domains, and technologies. 

\vspace{.3cm}
\noindent \textbf{How can event sourced systems be defined?}
 Section~\ref{sec:eventstoreandsystem} gives definitions of the different concepts in event sourcing and event sourced systems.
These definitions are based on our five years of experience in building an ESS, and they are augmented with the interviews.
The experiences of the interviewed engineers add nuance and variation options to the different concepts, making them reflect the view of practitioners. 
Concepts and codes extracted from the interviews scoped our definition: the engineers provided us the topics to define through the interviews. 

\vspace{.3cm}
\noindent \textbf{How can event sourced data structures be evolved?}
 Five event schema evolution techniques are discussed in Section~\ref{sec:evolution}: \emph{versioned events}, \emph{weak schema}, \emph{upcasters}, \emph{in-place transformation}, and \emph{copy-transform}.
For every technique the benefits and liabilities as discussed with the interviewed engineers, as summarized in Table~\ref{tab:techniques-summary}.
Almost all engineers have experience with multiple techniques, often combining them in a single system.
As all techniques have their benefits and their liabilities we did not found a single technique that would be applicable in all scenarios.
We conclude the section with general advice on when to apply specific techniques, and how to combine the techniques.

\vspace{.3cm}
\noindent \textbf{What are the challenges faced in applying applying event sourcing?}
Five challenges that the interviewed engineers experienced are discussed in Section~\ref{sec:challenges} and summarized in Table~\ref{tab:challenges}.
We address the steep learning curve in Section~\ref{sec:eventstoreandsystem} by giving definitions and operations that can be used in discussing and teaching of ESSs.
Evolution is discussed in detail in Section~\ref{sec:evolution}, again using the concepts and operations to explain and characterize the different techniques.
The other three challenges, lack of technology, rebuilding projections, and privacy, are presented as a start for a research roadmap.
We call for researchers to further explore these challenges.

The main scientific contributions are found in Sections~\ref{sec:research-approach} and~\ref{sec:challenges}. 
In the research approach, we aim to inspire future architecture researchers to use similar qualitative techniques, such as GT, for the explication of architecture knowledge from practitioners. 
Secondly, a set of research challenges is provided for software engineering researchers to challenge the knowledge around event sourcing in large software systems. 
Additionally, we are excited to define and document such an important software pattern for the scientific community. 

\balance

\section*{Acknowledgements}
The authors thank all the engineers for sharing their valuable experience and their willingness to contribute to this study. 
Furthermore, we would like to thank Paris Avgeriou, Fabiano Dalpiaz, Jurriaan Hage, André van der Hoek, John Mylopoulos, Alexander Serebrenik, Jan Martijn van der Werf, Greg Young, Uwe Zdun, and all the anonymous reviewers for their constructive feedback on earlier drafts.

\clearpage

%
%

\appendix
\section{Interview Protocol}
\label{sec:interviewprotocol}

\balance

\subsection*{Context related questions}

\begin{enumerate}
  \item Please introduce yourself, the company, the product, and your role in the development.
    \begin{enumerate}
      \item How many years is the system in production?
      \item How many installations are there of the system (single on-premise custom-made, single cloud SaaS, multiple on-premise customers, ...)?
      \item What is the load on the system in terms of users/ traffic (events?)? Can you give a rough estimate?
    \end{enumerate}
  \item Why is event sourcing applied in this software system? 
    \begin{enumerate}
      \item If this decision is already a few years old, is event sourcing still applicable of would the team decide otherwise with the current knowledge?
    \end{enumerate}
  \item What is the technology stack?
  \item Could you give a summary of the size of the system in terms of event sourcing? For instance in terms of different stream types, stream instances and number of events.
\end{enumerate}

\subsection*{Versioning related questions}

\begin{enumerate}
  \setcounter{enumi}{4}
  \item What strategy do you use for event versioning? (Elaborate on the why)
    \begin{enumerate}
      \item When using weak serialization: How do you deal with not being able to perform certain operations? Does it bother you, or not?
      \item When using upcasters: How many upcasters are there? What is the longest chain of upcasters? How do you manage them?
      \item When using in-place scripts: How do you validate the correctness? What about the audit log, how do you deal with re-writing?
      \item When using conversion: How long does it take? What about the audit log, how do you deal with re-writing?
    \end{enumerate}
  \item Do you need/ want the audit features? (What is the level of immutability?)
  \item What is your strategy for the query-side? How do you keep this in sync?
  \item How often are new versions released, and who performance the upgrade? 
  \item What kind of upgrade strategy is used? How do you deploy an upgrade? 
    \begin{enumerate}
      \item Do you have any SLAs based on the domain/product? (such as 24/7, 9 to 5)
    \end{enumerate}
\end{enumerate}

\subsection*{Other topics}

\begin{enumerate}
  \setcounter{enumi}{9}
  \item Do you use ProcessManagers/Sagas? Anything special for those?
  \item Are you satisfied with the current upgrade and versioning strategy? If not, what would you like to see differently?
  \item What do you see as future challenges of ESSs?
  \item Can you apply event sourcing without DDD?
  \item What would your approach be to building a huge system?
\end{enumerate}

\subsection*{Closing}

\begin{enumerate}
  \setcounter{enumi}{14}
  \item What did we miss? What should we have asked?
  \item With whom should we talk?
\end{enumerate}

\clearpage

\balance
\bibliographystyle{cas-model2-names}
\bibliography{main}   

\end{document}